\documentclass[10pt, aps, prc, superscriptaddress, nofootinbib, 
               amsmath, 
               twocolumn, preprintnumbers, 
               showpacs, 
               raggedbottom, 
               floatfix, 
               longbibliography, 
               colorlinks=true, linkcolor=blue, citecolor=blue, urlcolor=blue, 
              ]{revtex4-1}
\pdfoutput=1
\usepackage[T1]{fontenc}
\usepackage[utf8]{inputenc}
\usepackage[secnoindent, 
            ]{my_paper} 
\usepackage{my_acronyms}        
\usepackage[english]{babel}     
\usepackage{adjustbox}          
\usepackage{colortbl}
\usepackage{xparse}             
\usepackage[caption=false]{subfig}
\usepackage[percent]{overpic} 

\usepackage[
            tightpage]{preview}

\renewcommand{\vect}[1]{\boldsymbol{#1}}

\newcommand{\beq}{\begin{equation}}
\newcommand{\eeq}{\end{equation}}
\newcommand{\bea}{\begin{eqnarray}}
\newcommand{\eea}{\end{eqnarray}}

\begin{document}

\newlength{\mybaselineskip}
\setlength{\mybaselineskip}{\baselineskip}

\title{Reply to the Comment of S. Ayik and D. Lacroix, posted as arXiv:1909.1361v1~\cite{Ayik:2019},  on the recent article 
``Fission Dynamics  of $^{240}$Pu from Saddle-to-Scission and Beyond'' by 
Bulgac et al, published as Phys. Rev. C {\bf 100}, 034615 (2019)~\cite{Bulgac:2019b} }

\author{Aurel Bulgac}%
\email{bulgac@uw.edu}%
\affiliation{Department of Physics,  University of Washington, Seattle, Washington 98195--1560, USA}

\date{\today}

\begin{abstract}

A point-by-point answer to the comment authored by S. Ayik and D. Lacroix is presented. 
At this point in time this text is not aimed at being submitted to Phys. Rev. C or any other journal, unless 
the authors of the comment choose to follow such an avenue. I also suggest a possible formulation of a 
stochastic mean field approach free of the difficulties in the stochastic mean field model due to \textcite{Ayik:2008}.

\end{abstract}

\preprint{NT@UW-19-15}

\maketitle


\section{Rebuttal}

Ayik and Lacroix raise four points in their comment~\cite{Ayik:2019} and I shall address them in that order. 

\begin{enumerate}

\item The full sentence in Ref.~\cite{Bulgac:2019b} on page 21, which Ayik and Lacroix address is:

``In the stochastic mean field model 
\emph{fluctuations only stem from the fluctuations in the initial density}~\cite{Tanimura:2017}
and the time evolution is exactly the usual time-dependent mean field. 
This ad hoc assumption 
is at odds with the Langevin approach \textcolor{blue}{and also with the path-integral 
approach, in which fluctuations along the entire path are relevant.}''

but they chose to drop the text in blue, and thus they chose to ignore the thrust of the entire comment. 
The text in italics is a direct quote from \textcite{Tanimura:2017}.

Indeed, if for a given deterministic physical system with many degrees of freedom one chooses to 
suggest a description in terms of a reduced number of degrees of freedom, the dynamics of the reduced 
system can appear random, as is the case of a Brownian particle for example. 

The main problem with the stochastic mean field (SMF) model~\cite{Ayik:2008} (at least as applied to nuclei in the context 
of Ref.~\cite{Tanimura:2017}) is that is not the situation discussed by Ayik and Lacroix in this point 1, because:

{\it In the mean field approximation for an isolated nucleus all single-particle degrees of freedom 
are active and there are no ignored, bath, or environmental degrees of freedom}.  

As a matter of fact in 
mean field approaches there are too many degrees of freedom, while strictly speaking one should include 
only the intrinsic degrees of freedom. For an ergodic system obviously the memory 
of the initial configuration
does not matter.
In the case of  Brownian motion, if one were to include all atomic
degrees of freedom, either at a mean field level or even exactly, along 
with those of the Brownian particle, the motion of the 
Brownian particle would be totally deterministic and not random. 
The same would apply to an open quantum system if all degrees of 
freedom (including the environment) would be included in the description. A nucleus is not an open system, 
unless one couples it to the vacuum fluctuations of the electromagnetic field, 
to the surrounding electrons, or to weak interactions.

 As generations of theoretical physicists know very well, the equation for the 
 Brownian motion is ``derived'' only by making many assumptions, 
 which sometimes are accurate enough for practical purposes and sometimes are 
 not, as is the case of fractional Brownian motion or when memory effects are relevant.
 The Langevin or the Fokker-Planck equations have limits of validity, 
 which are not always clearly understood.
 The Markovian character is a particular limit and the absence of 
 memory effects is not always a correct assumption, 
 and the noise is not always Gaussian~\cite{Kampen:1990,Montroll:1987}.

In the path integral approach for example fluctuations are present at all times. In a treatment
in which only a reduced number of degrees of freedom are explicitly included, random fluctuations will appear 
at all times due to different reasons, and only one source of the apparent stochasticity could 
be traced back to the ignorance of the initial conditions. Chaoticity, ergodicity, mixing behavior and other related phenomena 
are ubiquitous in Nature and not because the initial conditions are not known.
In Ref.~\cite{Tanimura:2017} the authors do not discuss the evolution of a reduced set of collective 
degrees of freedom. Again, even if they choose to do that, they should remember 
that the source of randomness is not due to the ignorance of the initial conditions alone.
In the SMF model the initial conditions are the only source of stochasticity. 
Such a statement is indeed correct only in 
the case of integrable models (such as ideal gases in special types of enclosures), 
basically for models with vanishing Lyapunov exponents. Even an ideal gas 
is chaotic in most types of enclosures. 

\item In Ref.~\cite{Bulgac:2019b} it was not claimed that interpreting 
a phenomenological approach event-by-event 
makes sense. One important claim was that the properties of the 
``stochastic events'' in the SMF model
lead  to unphysical results after ensemble averaging.

\textcite{Ayik:2019} write in their note: 

``The statistical properties of the initial densities are 
obtained by imposing that the average values of first and second moments equals the quantum 
average. To be able to match statistical and quantal average, {\it there is no other choice} than 
exploring a wider class of one-body densities usually allowed for Fermi systems.'' 

(The emphasis/italics in their quote is mine.) The authors do 
not quantify what ``allowed for Fermi systems'' really means and
the conclusion that there is no other choice is false.
Again, in Nature not all processes are Gaussian. And there  
are other choices if the model leads to unphysical outcomes! 

{\it Moreover, since fluctuations are real and exist 
in the ground state of any interacting quantum system, upon introducing the ground state fluctuations
one should obtain the ground state properties,  and not 
the properties of an arbitrarily chosen excited state, which is what is done in SMF model, in particular
by \textcite{Tanimura:2017}.} 

According to the email correspondence between myself and 
the authors of Ref.~\cite{Tanimura:2017}, the variance of 
the single-particle occupation probabilities, see Eq. \eqref{eq:sigma}, 
were chosen from the pairing correlations of the 
constrained mean field nuclear configuration at zero temperature, not from the fluctuations of 
the target excited state they discuss. 
In Density Functional Theory the fluctuations, known as correlations or 
corrections on top of mean field, when included properly, would  
lead to a better estimate of the ground state energy, 
not the energy of an excited state. 
\footnote{The Density Functional Theory DFT is a guiding 
approach adopted in nuclear physics as well.  
I use ``guiding approach adopted in nuclear physics''  
as  there exist somewhat different opinions in  nuclear literature.}
See more about this aspect also below.

There is a major difference between an exact approach, such as the path integral approach, Eq. (1)
 in Ref.~\cite{Bulgac:2019b} and the SMF model. All unphysical contributions to a path integral cancel exactly, 
 but that is not the case in the SMF model.  While the average particle number is correct, the variance of the 
 particle number  does not vanish in SMF model (except in the strict independent particle 
 limit at zero temperature,  when fluctuations are absent),
 \beq
  \overline { \left ( N -\overline{N}\right  )^2  }=\sum_{k}n_k(1-n_k),
  \eeq 
 where 
 \bea
 &&N =\Tr \rho= \sum_k (n_k+\xi_{kk}),\quad \overline {N} = \sum_k n_k,\\
 && \rho(x,y) = \sum_k  n_k\phi_k(x)\phi_k^*(y) + \sum_{k,l} \xi_{kl} \phi_k(x)\phi_l^*(y), \label{eq:nnn}\\
 && \rho(x,y)=\rho^*(y,x)= \int \!\!\!d\mathrm{N} \; c(\mathrm{N})\rho(x,y,\mathrm{N}), \label{eq:N}
  \eea
 where $\rho(x,y)$ is the (Hermitian) stochastic single-particle density matrix, 
 $\xi_{kl}$ are time-independent, independent  Gaussian complex random numbers with zero mean
  and variance
\bea
\sigma^2_{kl} 
= \overline{\xi^{\lambda}_{kl}\xi^{\lambda *}_{kl}} 
= \frac{1}{2}\left [ n_k(1-n_l) + n_l(1-n_k) \right ], 
\label{eq:sigma} 
\eea
 $n_k$ are single-particle occupation probabilities, $\phi_k$ are orthogonal single-particle 
 wave functions $\langle \phi_k|\phi_l\rangle =\delta_{kl}$, and
 the overline stands for the statistical ensemble average. In Eq. \eqref{eq:N} I used the notation 
 $\rho(x,y,\mathrm{N})$ for the projected single-particle density matrix on a fixed particle 
 number $\mathrm{N}$ (note the font for $\mathrm{N}$, which is different from the one for $N$), thus 
 \beq
 \int\!\!\!dx \; \rho(x,x,\mathrm{N}) \equiv \mathrm{N}.
 \eeq
 
 In the SMF model $N$ can acquire any real value, in contradistinction to the Hartree-Fock-Bogoliubov 
 approximation where $N\ge 0$ is an even integer (for an even-even nucleus). 
  
 In their point 3, to which I shall return below, \textcite{Ayik:2019}
 claim that there is a need for a coarse-graining procedure  of the density matrix 
 $\rho(x,y)$ (still not yet mentioned as such in the literature dealing with the SMF model).
 I am guessing that this coarse-graining procedure is somewhat similar 
 to the Husimi averaging procedure of the Wigner distribution.
 I do not think such a coarse-graining of the number density  is a good solution.
  
 One can naively counteract the statement 
 that the particle number fluctuates in SMF model  with the counterargument that at 
 finite temperatures in mean field models particle number fluctuates, 
 or the same is true in mean field models with 
 pairing correlations taken into account. 
 
 First, a fissioning nucleus is not a system at a finite temperature, 
 it is an isolated system and there is no thermal bath, no environment, and no 
 ignored degrees of freedom in the description discussed in 
 Refs.~\cite{Tanimura:2017,Bulgac:2019b,Ayik:2019}. 
 Introducing a temperature is a pure phenomenological 
 approach and {\it there is no microscopic recipe} on how to consistently define a 
 temperature or to claim that the system attained any kind of meaningful equilibrium 
 with a well defined temperature during the descent from saddle-to-scission.
 One can make assumptions of course, and that is what one does 
 in phenomenological approaches or models, 
 and one might even get lucky,
 but that is not a theoretical argument.  
 
 Second, in the case of pairing correlations there is 
 always the possibility to project onto the correct particle
 numbers, if that is indeed needed for a more precise description of the observables. 
 Quantum mechanical systems can also
 be described in the canonical ensemble at finite temperatures~\cite{Rossignoli:1994,Fanto:2017}
 and not only in the grand canonical ensemble 
(which is technically more convenient). There was no recipe defined in the SMF 
 model  so far on how to project the correct particle number, 
 when  $N$ is  a real number.
  
 If the variance of 
 the particle number for the entire fissioning nucleus
 is incorrect in simulations, how can \textcite{Tanimura:2017} make the case
 that the emerging variances of the proton and neutron numbers of 
 the fission fragments have anything to do with reality? 
 
 And a related very important aspect, if the particle number fluctuates, which 
 density matrix should one use in calculating the energy of a nucleus. 
 Should one use $ \rho(x,y,\mathrm{N})$? 
 Or the unprojected particle number density matrix 
 $\rho(x,y)=\int \!\!d\mathrm{N} \; c(\mathrm{N})\rho(x,y,\mathrm{N})$  and perform the particle 
 projection of the total energy only afterwards? 
Using the unprojected density matrix $\rho(x,y)$ would lead to unphysical contributions to the total energy
 from configurations with the wrong particle number. 
 Unlike in the case of spontaneous symmetry breaking, one cannot make the case that performing 
 a particle  projection after the total energy is computed with the unprojected density matrix $\rho(x,y)$ 
 would be a valid or accurate procedure. One can make a fair guess that 
 \textcite{Tanimura:2017} computed the nucleus
 energy with $\rho(x,y)$ and that no particle projection was performed. I discuss this aspect
 also below from a slightly different prospective.

 The same applies as well for the 
 excitation energies of the fission fragments distributions, upon 
 I touch below too.

\item In their point 3, Ayik and Lacroix argue that Eq. (D16) in Ref.~\cite{Bulgac:2019b} is 
interpreted incorrectly physically.

In this case Ayik and Lacroix mix two issues, namely the difference between the quantum operators
\bea
\hat{O}_1(x,y)&=& \psi^\dagger(x)\psi^\dagger(y)\psi(y)\psi(x), \label{eq:eq3} \\
\hat{O}_2(x,y)&=&\rho(x)\rho(y) =\psi^\dagger(x)\psi(x)\psi^\dagger(y)\psi(y)\nonumber \\
&=& \hat{O}_1(x,y)+ \delta(x-y)\rho(x). 
\eea
Their argument is that the fluctuations of the density-density operator $\hat{O}_2(x,y)=\rho(x)\rho(y)$ 
diverge when the two spatial points coincide (obviously), which is indeed a correct statement~\cite{LL5:1980}. 
However, the energy density functional is obtained from the expectation value of the operator 
$\hat{O}_1(x,y)$, which has no divergence. There 
is a good reason why in many-body theory 
the operators in second quantization are normal ordered.

Concerning the parameter free statement of the SMF model made often in either 
discussions or in print by Ayik and collaborators, a 
quick look at the Fig. 1 from the Supplemental Material of Ref.~\cite{Tanimura:2017} 
will reveal just the opposite. For a quadrupole deformation $Q_{20}\approx 160$ (barn)
(the choice of which value is treated as a free parameter), 
depending on the width of the single-particle window the authors choose (which is another 
free parameter) they obtain two different total energies of the nucleus. 
And arguably, one could have chosen any other single-particle energy window for other purposes.
Thus the end justifies the means.

One could choose which single-particle occupation 
probabilities  fluctuate in a many other ways too. Since the number of protons and 
neutrons are not equal one can make the case that the
number of proton and neutron levels where fluctuations are allowed are different also, 
e.g. in a ratio proportional to Z/N? Why 
allow for relatively more protons occupation probabilities to fluctuate than for neutrons? 
And that is of course only one of many possible arbitrary choices.
  
In their comment \textcite{Ayik:2019} make the claim that 
one should coarse-grain the local density fluctuations  
in the energy density functional over spatial regions containing at least one particle. 

Let us see how this suggestion might work. The average volume 
occupied by a single nucleon in a nucleus is $\approx 6$ {fm}$^3$. In numerical simulations as those 
described in Refs.~\cite{Bulgac:2019b,Tanimura:2017,Bulgac:2016}. 
The densities are already coarse-grained over a volume 
$l^3\approx 0.5\ldots 2$ {fm}$^3$ (depending on the specific simulation), 
where $l$ is the lattice constant. The value of the lattice constant used in 
Refs.~\cite{Bulgac:2016,Bulgac:2019b} corresponds to a single-particle momentum cutoff 
$p_c=\tfrac{\hbar\pi}{l}\approx 500$ {MeV/c} and $l^3\approx 2$ fm$^3$.
This value of the momentum cutoff is basically the value used in modern
chiral EFT models of the nucleon-nucleon interactions, which can 
be used in the construction of quite accurate nuclear energy density 
functionals~\cite{Navarro-Perez:2018,Melendez:2019,Binder:2018}.  
Applying the arguments of the Appendix D in Ref.~\cite{Bulgac:2019b}, 
in particular Eq. (D17) following from Eq. (D16), 
one would obtain that the expected value of a typical term in 
the energy density functional is $N \tfrac{N}{V}$
\beq
\int \!\!\!d{\bm r} \;\rho^2({\bm r},{\bm r},t) 
 \approx N \frac{N}{V}\ll  \frac{(gN_{xyz}- N) N }{2V}, \label{eq:eq4}
\eeq
while in the SMF model this value is orders of magnitude larger, namely $\tfrac{(gN_{xyz}- N) N }{2V}$.
Here $N_{xyz}$ are the number of lattice points in the simulation 
box and $g=4$ is the spin-isospin degeneracy.  Also, if adopting \textcite{Ayik:2019} 
coarse-graining procedure $N_{xyz}\approx 27,000$ could be equally interpreted as the number of spatial cells
over which the coarse-graining is performed. 
Even by increasing the lattice constant to 2 fm and thus decreasing to $N_{xyz}\approx 6,700$, 
and achieving a coarse-graining volume of 8 fm$^3$, 
which is greater than the average value occupied by a single nucleon, the actual
size of the  fluctuations of the term \eqref{eq:eq4} in SMF model is still enormous. 

\textcite{Ayik:2019} state that:

``the energy obtained by averaging the Hartree-Fock energy over events will 
match the energy of the initial state and no divergence will occur.  
However, only when SMF approach is applied 
to a density functional theory (DFT), the special attention of the terms like 
the one discussed in Eq. (D16) should be made.'' 

When reading the text by \textcite{Tanimura:2017} 
I have not been able to find any place where a special attention seems to have been paid
for the calculated energy of either of the fissioning nucleus or of the energies of the fission fragments. 
Was this kind of special attention paid in any of all previous published calculations using the SMF model?

Maybe the above quote has to be interpreted in the following manner: 

The SMF model does not 
lead to divergencies only in the strict Hartree-Fock approximation with density independent interactions. 
The popular Skyrme interaction depends on the density.  

I suspect that in this case
the authors suggest that in the interaction one should use the ensemble averaged density and as 
result the Hartree-Fock expectation of this type of ensemble averaged density dependent interaction contains 
only terms quadratic in the density, and that is why they claim there are no divergencies. Well, I am not sure 
that is correct either, as
\bea
&&\overline{ \rho(x,x)\rho(y,y) } = \sum_k n_k|\phi_k(x)|^2 \sum_l n_l |\phi_l(y) |^2  \label{eq:n1}\\
       &+&\sum_{kl} \frac{1}{2}(n_k+n_l-2n_kn_l) \phi_k(x)\phi_l^*(x)\phi_l(y)\phi_k^*(y) \nonumber
\eea
and
\beq
\overline{ \rho(x,x)\rho(x,x) } = \sum_k  n_k|\phi_k(x)|^2\sum_l |\phi_l(x)|^2 \label{eq:n2} 
\eeq
still looks diverging in the absence of an upper limit in the sum over $l$. 

\textcite{Ayik:2019} state further:

``The truncation of the particle-hole space in a narrow energy
range around Fermi surface provides a possible for
the coarse-graining of the local density fluctuations.'' 

and indeed, if the sums have an upper limit there is no divergence, but the values of these sums increase
with increasing the upper limit and one still obtains unphysical energy estimates, 
see also Eq. \eqref{eq:eq4} and the ensuing discussion.

There is another unclear aspect concerning the coarse-graining suggestion made by \textcite{Ayik:2019}. 
In the Supplemental Material of Ref.~\cite{Tanimura:2017} for the same nuclear shape 
with quadrupole momentum $Q_{20}\approx$ 160 barn the authors use two 
coarse-graining procedures and obtain two different excitation energies.  
Thus one can imply that this suggested 
coarse-graining procedure is directly related with the desired target excitation energy of the nucleus. 
The wider the energy window in the single-particle 
levels corresponds to a higher total nucleus energy. Shall we interpret this as the additional prescription
not made clear until now in the SMF model, that is that the excitation energy 
determines the scale of the coarse-graining procedure, needed to avoid divergencies? 

One could have opted 
to introduce the fluctuations at $Q_{20}\approx$ 100 barn,
where the nucleus just emerged from under the barrier. 
The descent towards scission 
is without argument  a non-equilibrium one. Why then not try to describe the 
entire process in a truly dynamic manner?
At this deformation the nucleus energy is the 
initial energy, and thus no coarse-graining would be needed.  

At $Q_{20}\approx $ 100 barn there is no stochasticity in the single-particle density matrix. 
According to the philosophy of the SMF model applied to fission by \textcite{Tanimura:2017}   
when evaluating various observables, while evolving from the configuration
where fluctuations are not ``needed'' (at $Q_{20}\approx$ 100 barn), 
towards more elongated nuclear shapes, one should use ensemble averages 
of  observables expressed through a stochastic single-particle density matrix. The stochasticity should be 
implemented in the initial density matrix at some time before the measurement is performed in the final state.
This initial state, where there are no fluctuations ``needed,'' represents a very unique kind 
of bifurcation configuration for nuclear trajectories, whose ensembles are 
supposed to describe correctly fission dynamics in particular. Moreover, each member of this ensemble
of single-particle density matrices has a fully deterministic time evolution.

Thus the state with $Q_{20}\approx $ 100 barn is in some sense a highly 
metastable state (for the lack of a better term). 
However, any other initial point on the outer potential energy surface, with an energy higher than 
the ground state energy of the nucleus can be 
reached by exciting the nucleus, e.g. with $\gamma$-rays and subsequent 
barrier tunneling. Therefore all these 
points on the outer potential energy surface  are also highly metastable states 
as well, as they have no initial fluctuations according to the SMF model.
One can then easily make the argument that the entire potential energy surface 
is a manifold of such metastable states. 

According to the philosophy of the 
SMF model applied to fission, the measure of the stochasticity of the single-particle density 
matrix is determined only by the energy difference between the initial energy and the energy of the arbitrarily
chosen target configuration on the potential energy surface and, hopefully, one should obtain the same final 
configuration after scission irrespective of where one starts implementing the stochasticity of the density matrix.
\textcite{Tanimura:2017} results 
however do not support such a strong conclusion. The SMF model works (almost) correctly, but
not always, as one has to choose the initial state according to some murky procedure, basically when it works.

In my opinion the coarse-graining ``solution'' suggested by \textcite{Ayik:2019} in 
their point 3 is full of inconsistencies, it is an {\it ad hoc} prescription with little if any microscopic  underpinning. 
Notably, \textcite{Tanimura:2017} never mention this coarse-graining procedure, 
and it was never explicitly mentioned in any other applications of the SMF model. Of course, 
this coarse-graining procedure also leads to all the difficulties discussed in my rebuttal of the points 1 and 2.

\item Ayik and Lacroix dispute the veracity of our statement on page 21 of Ref.~\cite{Bulgac:2019b}:

``Since in the stochastic mean field method \emph{fluctuations only stem from the
fluctuations in the initial density}~\cite{Tanimura:2017} one 
would expect that their conclusions should parallel ours, 
as we have considered  a relatively large set of initial conditions with a similar spread in initial energies 
and deformations. ''

\textcite{Ayik:2019} make the argument that while we have
included only axially symmetric initial conditions with various values of 
the multipole moments $Q_{20}$ and $Q_{30}$, we should have also considered  at least 
non-axially deformed initial states as well, which play a crucial role 
(unlike the axially symmetric states) according to them. 

Indeed, in Ref.~\cite{Bulgac:2019b} we have exemplified 
our conclusions with only axially deformed initial conditions. However, we have also mentioned in 
passing that considering much more complex fluctuations (and not only in the initial conditions, 
see Ref.~\cite{Bulgac:2019a}) the fission dynamics had a very similar qualitative behavior. Apart from 
that we have been informed for quite some time by Piotr Magierski of simulations he performed 
(unfortunately unpublished), considering also non-axially deformed initial states with results 
basically paralleling our results. Due to the very strong damping of the collective motion during the
descend of the nucleus from saddle-to-scission (which has been established for the first time 
in a fully microscopic treatment in Ref.~\cite{Bulgac:2019b}) 
the memory of the initial conditions is rather quickly forgotten, 
see in particular Fig. 9 in Ref.~\cite{Bulgac:2019b}. In our experience so far we see no reasons to 
concur with Ayik and Lacroix's statement made in their comment only, that the 

``absence of 
restriction (to initial axially symmetric configurations) in SMF turns out to be 
crucial to grasp beyond mean field effects.''~\cite{Ayik:2019}  
 
\end{enumerate}

\section{An Alternate  Stochastic Mean Field Model}

Adding stochasticity to the mean field is likely an idea worth pursuing.  Here I will suggest an alternative extension of Ayik's 
SMF model~\cite{Ayik:2008}, which is free of a range of deficiencies discussed in Ref.~\cite{Bulgac:2019b} and above. 
Let us consider a generic type of density matrices:
\bea
&&\!\!\!\!\!\!\!\rho=\frac{1}{Z}\exp \left  [ -\beta \sum_{kl} (\varepsilon_{kl}a^\dagger_ka_l +
\Delta_{kl}a_k^\dagger a_l^\dagger +\Delta^*_{kl} a_l a_k -\mu)  \right ],\\
&&\Tr \rho =1, \label{eq:n5}
\eea
where $\varepsilon_{kl}$ and $\Delta_{kl}$ are random (Gaussian) numbers, with variances chosen appropriately.

Since a physically acceptable density matrix is positive definite and Hermitian it can be diagonalized 
simultaneously with mean field Hamiltonian and then the stochastic density matrix acquires the form:
\bea
\rho &=& \frac{1}{Z}\exp \left  [ -\beta \sum_k (\varepsilon_k+\delta \varepsilon_k-\mu) a^\dagger_ka_k \right ],\\
Z&=&\prod_k\{1+\exp [-\beta(\varepsilon_k+ \delta \varepsilon_k-\mu)]\},\\
n_k&=& \frac{1}{\exp[\beta(\varepsilon_k+ \delta \varepsilon_k-\mu)]+1},\\
N&=&\sum_k n_k, \label{eq:mu}
\eea
where $n_k$ are single-particle occupation probabilities.
Eq. \eqref{eq:mu} fixes the chemical potential for each realization of the stochastic density matrix 
and $\beta$ should be chosen according to the needs. e.g. to fix the excitation energy.
Since the fluctuations of the energy levels around the average spectrum are typically of the 
Generalized Orthogonal Ensemble (GOE) type in even-even 
nuclei or even simple systems, such as quantum billiards~\cite{Brody:1981,Bohigas:1984,Mehta:1991,Brack:1997,Alt:1995,Mitchell:2010},  
it makes sense to chose 
$\delta \varepsilon_k$ from such an ensemble. $\varepsilon_k$ could be chosen from an 
averaged single-particle spectrum following Strutinsky's or the quantum chaos 
prescriptions~\cite{Strutinsky:1967,BRACK:1972,Brack:1997}
adding on top $\delta \varepsilon_k$ according to the GOE prescription.
Often instead of the GOE many consider two-body random ensembles~\cite{French:1970,Wong:1972,Papenbrock:2007}
or banded random matrices, which lead to correct description of the average many-body 
level density~\cite{Brink:1979} and also to their non-equilibrium dynamics \cite{Agassi:1975,Brink:1979,Bulgac:1995,Bulgac:1996,Bulgac:1998}. 
By maintaining the Markovian character of the non-equilibrium dynamics of a many-fermion system,
but renouncing the Gaussian character of the fluctuations one can significantly enlarge the quantum 
evolution types to fractional kinetics~\cite{Kusnezov:1999}.  These suggestions should be taken with a grain of 
salt at this point, as some of them I did not have a chance yet to carefully evaluate.

When necessary, one can perform a 
particle projection as well~\cite{Rossignoli:1994,Fanto:2017}. 

\begin{acknowledgments}

This work was supported by U.S. Department of Energy,
Office of Science, Grant No. DE-FG02-97ER41014 and in part by NNSA
cooperative agreement DE-NA0003841.     

I thank L.M. Troy, K.J. Roche, and I. Stetcu for reading a draft and suggesting several improvements 
of the text.

\end{acknowledgments}



\providecommand{\selectlanguage}[1]{}
\renewcommand{\selectlanguage}[1]{}

\bibliography{../local_fission}

\end{document}